# Characterization of relativistic electron bunch duration and travelling wave structure phase velocity based on momentum spectra measurements on the ARES linac at DESY

T. Vinatier[a,1], R. W. Assmann[a], C. Bruni[b], F. Burkart[a], H. Dinter[a], S. M. Jaster-Merz[a], M. Kellermeier[a], W. Kuropka[a], F. Mayet[a] and B. Stacey[a]

[a]Deutsches Elektronen-Synchrotron DESY, Notkestr. 85, 22607 Hamburg, Germany
[b]IJCLab, Université Paris Saclay, Bâtiment 100, 15 rue Georges Clémenceau, 91405 Orsay, France
[1]Corresponding author: thomas.vinatier@desy.de

*Abstract*

The ARES linac at DESY aims to generate and characterize ultrashort electron bunches (fs to sub-fs duration) with high momentum and arrival time stability for the purpose of applications related to accelerator R&D, e.g. development of advanced and compact diagnostics and accelerating structures, test of new accelerator components, medical applications studies, machine learning, etc. During its commissioning phase, the bunch duration characterization of the electron bunches generated at ARES has been performed with an RF-phasing technique relying on momentum spectra measurements, using only common accelerator elements (RF accelerating structures and magnetic spectrometers). The sensitivity of the method allowed highlighting different response times for Mo and $Cs_2Te$ cathodes. The measured electron bunch duration in a wide range of machine parameters shows excellent agreement overall with the simulation predictions, thus demonstrating a very good understanding of the ARES operation on the bunch duration aspect. The importance of a precise in-situ experimental determination of the phase velocity of the first travelling wave accelerating structure after the electron source, for which we propose a simple new beam-based method precise down to sub-permille variation respective to the speed of light in vacuum, is emphasized for this purpose. A minimum bunch duration of 20 fs rms, resolution-limited by the space charge forces, is reported. This is, to the best of our knowledge, around 4 times shorter than what has been previously experimentally demonstrated based on RF-phasing techniques with a single RF structure. The present study constitutes a strong basis for future time characterization down to the sub-fs level at ARES, using dedicated X-band transverse deflecting structures.

## 1. Introduction

A large number of scientific applications require ultrashort electron bunches with a duration typically below 100 fs rms and ideally even reaching the (sub)-fs level, e.g. ultrafast radiation pulse generation through free electron lasers [1] or wakefields [2], ultrafast imaging via diffraction and microscopy [3,4], ultrafast pulse radiolysis [5], etc. The ARES linac at DESY [6,7] aims to generate and characterize such bunches, with high momentum and time arrival stability, for the purpose of applications related to accelerator R&D, e.g. development of advanced and compact diagnostics [8,9] and accelerating structures [10], test of new accelerator components, medical applications studies, etc. One of the main challenges to be faced towards this goal is to obtain a proper time characterization (duration, time profile, arrival time jitter) of these bunches.

To reach the required resolution, several types of dedicated diagnostics are currently used worldwide. A first type of method is based on electro-optical sampling of the electron bunch electromagnetic field with an ultrashort laser pulse [11]. This can be made single-shot if a chirped laser pulse is used [12,13]. However, the resolution is by definition limited to the duration of the probing laser pulse, which is currently around a few tens of fs rms for commercially available lasers, and therefore does not reach the single-digit fs to sub-fs level. Another general type of diagnostics is based on the bunch time profile reconstruction through a measurement of the frequency spectrum it emits in special conditions (e.g. Smith-Purcell radiation [14], coherent transition radiation [15], etc.). These methods can reach a resolution on the single-digit femtosecond level for the bunch duration [16], but require assumptions on the shape of the electron bunch time profile to be able to reconstruct it and to deal with the fact that only part of the radiation spectrum is usually recorded [17]. A last general type of diagnostics is the transverse deflecting structure (TDS) [18]. By using a downstream dipole magnet with a dispersive direction perpendicular to the streaking direction of the TDS, the entire bunch longitudinal phase-space (time vs momentum) can be recorded. TDS working in the S-band frequency range are nowadays widely used with resolutions down to tens of fs rms [19], and TDS at higher frequencies, especially in the X-band range, are in development and allow reaching a (sub)-fs resolution [20]. Advanced streaking schemes are also currently under investigation to reach sub-fs resolutions, like the use of THz pulses [21] or methods aiming to combine the streaking provided by a laser modulator on one transverse direction with the streaking provided by a TDS on the orthogonal transverse direction [22].

The aforementioned techniques all require additional cost and a dedicated space to be implemented. Furthermore, the installation of an adequate environment (vacuum system, RF system, laser transport line, detectors, etc.) is also required for operation. Finally, for some of these techniques, the electron bunch has to interact in a controlled and synchronized way with an external radiation pulse (e.g. laser) introducing further complexity.

Several methods (thereafter named RF-phasing methods) exist to measure the bunch duration and time profile by using only common elements virtually present on all research electron accelerators (RF accelerating structures, spectrometer and imaging screens), and therefore do not require additional cost and space, e.g. zero-phasing technique [23], phase-scan methods [24-27], time-dependent transverse field components of $TM_{01}$ mode [28], longitudinal phase-space tomography [29,30]. Despite not reaching the single-digit femtosecond resolution, these methods are still attractive to be used during the commissioning phase of an accelerator, before the implementation of dedicated diagnostics, but also on accelerators where no dedicated diagnostics are planned to be installed. This is for example the case on small accelerators for cost and/or space reasons or on the injectors for synchrotron light sources where the bunch duration is not a key parameter but can be of interest to be measured.

The primary diagnostics intended at ARES to diagnose the ultrashort electron bunches (see Table 1) are two PolariX X-band transverse deflecting structures [31-33], which are the product of a collaboration between CERN, DESY and PSI, for which the commissioning phase is expected to start in the second half of 2023. In addition to a sub-fs resolution, it has the feature of a variable streaking direction, thus allowing advanced tomographic reconstruction of the bunch distribution [8,33].

In this paper, the commissioning phase leading to the first characterization of the duration of the electron bunches generated by the ARES linac at DESY (see Sec. 2), which produced its first beam end of 2019 [6], is presented. The characterization of the bunch duration (see Sec. 5) relies on the use of a phase-scan method [24-25], which is based on beam momentum spectra measurements (see Sec. 3). A detailed comparison with the predictions from ASTRA simulations [34], a reference and well-benchmarked beam dynamics simulation code, is provided. An important requirement for this is a precise determination of the phase velocity of the first accelerating structure after the electron source, where the bunches are still not fully relativistic. To this aim, we propose a simple new method based on the measurement of the phase gap between momentum minima and maxima at the exit of this structure (see Sec. 4).

## 2. The ARES linac at DESY

The ARES (Accelerator Research Experiment at Sinbad) linac at DESY (see Fig. 1) is an approximately 45 m long linac operating in the S-band frequency range at 2.99792 GHz [6,7]. After the S-band gun, driven by a UV laser pulse at 257 nm, two 4.092 m long travelling wave accelerating structures (TWS1 and TWS2) operating in the $2\pi/3$ $TM_{01}$ mode bring the electron bunches to their final momentum, around 155 MeV/c at maximum (gun + 2 TWS operated on-crest). Several options are available to compress the electron bunches in time [35]: velocity bunching [36] in TWS1, magnetic compression in the bunch compressor [37] and a hybrid compression mode mixing velocity bunching and magnetic compression [38]. Two spectrometers (one after the gun and one at the end of the beamline) are available to diagnose the electron bunch momentum spectrum. The bunch charge is measured with a Faraday Cup located before TWS1, unless otherwise stated in the paper. Multiple screens, steerers and quadrupoles are located all along the beamline (not displayed in Fig. 1) for the purpose of beam transport, focusing and diagnostics. The target bunch properties of the ARES linac are summarized in Table 1. Some parameters are fixed for all the experimental results and simulations shown in the rest of the paper. They can be found in Table 2.

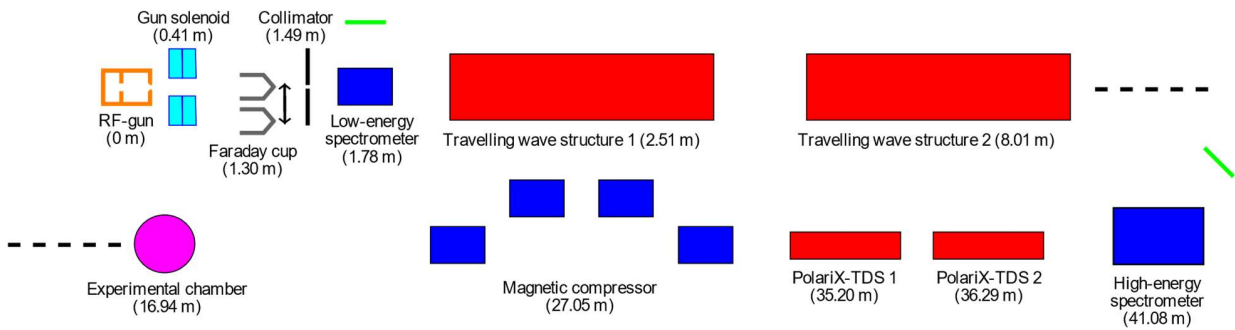

**Figure 1:** Schematic layout of the ARES linac at DESY [6,7] with the relevant elements for the study presented in this paper. The positions are given relative to the RF-gun cathode plane and at the entrance of each element.

It is important to note that there are no accelerating structures located downstream of the magnetic bunch compressor at ARES. It is therefore not possible to perform a bunch duration measurement of the bunches coming out of it via RF-phasing methods. For this purpose, the PolariX transverse deflecting structures will be needed. The RF-phasing methods are, however, applicable to measure the bunches coming out of TWS1, eventually compressed in time via velocity bunching, since TWS2 is located right downstream.

Table 1: Target electron bunch properties of the ARES linac at DESY

| Properties | Target value |
|---|---|
| Charge | 0.01 – 200 pC |
| Momentum | 20 – 155 MeV/c |
| Minimum momentum spread | 0.01% rms |
| Bunch duration | Sub-fs to ≈ 10 fs rms |
| Transverse emittance (norm.) | < 0.8 $\pi$.mm.mrad rms |
| Arrival time jitter | < 10 fs rms |

Table 2: Fixed operation parameters of the ARES linac at DESY

| Parameters | Values |
|---|---|
| Rms duration of the laser pulse driving the RF-gun | 175 fs (Gaussian) |
| Transverse profile of the laser pulse driving the RF-gun | Uniform |
| TWS1 max. average gradient | 18.199 MV/m |
| TWS2 max. average gradient | 17.092 MV/m |

**3. The phase-scan method**

The bunch duration measurements presented in Sec. 5 are performed with a phase-scan method [24,25], introduced in Sec. 3.1. This method has been selected rather than the other RF-phasing techniques because of its high flexibility and large range of applications.

Indeed, using the time-dependent transverse field components of the $TM_{01}$ mode in a TWS, as proposed in [28], is limited to momenta of a few MeV/c. It is therefore not suited for ARES where the momentum is already ≥ 3.5 MeV/c after the gun and up to 80 MeV/c after TWS1. The phase-scan method is, on the opposite, also applicable at ultra-relativistic momenta.

The zero-phasing technique, as used in [23], has the advantage to maximize the time resolution, since the steepest slope of the RF accelerating field is used, thus maximizing the induced momentum spread. However, this comes with the drawback that the beam does not gain momentum in the RF structure, or only in a very limited amount. This lower momentum results in stronger space-charge forces [39] during the beam transport downstream of the RF structure. This can influence the beam momentum spread and therefore the duration measured via the zero-phasing technique. Moreover, the zero-phasing technique requires assumptions to be placed on the input bunch momentum spread (if the two zero-crossing phases are used) and also on the input bunch time-momentum correlation (if only one zero-crossing phase is used). Typically, it is assumed that the input bunch momentum spread is negligible compared to the one imparted by the structure at the zero-crossing phase, and eventually that the effect of the time-momentum correlation can be neglected. Contrary to this, the phase-scan method requires only limited assumptions on the longitudinal phase-space (see Sec. 3.1) and can be used with any phases of injection into the RF accelerating structure, also very close to on-crest where the momentum gain is maximal.

The phase-scan method gives less information than the tomography, which provides the bunch time profile and even the entire longitudinal phase-space. Nevertheless, the former is used on ARES for two practical reasons. First, it is much less sensitive to noise and jitter than the tomography. They are indeed washed out by the fact that the phase-scan method only requires the statistical moment (momentum spread) as input, while the tomography requires the full momentum spectrum. Second, the tomography requires a large number of input spectra (typically > 30) to be accurate and the phase-space reconstruction is computationally expensive (typically 1 to 30 minutes depending on the number of input spectra and the bin size of the reconstructed phase-space), while the phase-scan method already gives accurate results with less input spectra (around 10) and requires a negligible computing time for reconstruction (a few seconds). It is therefore better suited for live use in operation and to perform quick measurements.

*3.1. The underlying model*

The phase-scan method allows determining the statistical properties of the bunch longitudinal phase-space (rms duration $\sigma_t$, rms momentum spread $\sigma_p$ and rms time/momentum correlation $\sigma_{pt}$) through the measurement of its momentum spread with a magnetic spectrometer for at least 3 conditions of operation of one or several upstream-located RF accelerating structures. The determined properties are then the ones at the entrance of the first accelerating structure which operation conditions are varied. It is the analogon of the quadrupole scan technique in the transverse phase-space [40].

It is in principle possible to vary either the field amplitude and/or the bunch injection phase into the accelerating structures. However, it is practically easier to vary the phase and only this option is considered in this paper. Also, only the case of a single travelling wave accelerating structure (TWS2) is considered in this paper.

In order to establish the analytical model on which the phase-scan method relies, several assumptions are made:

1. The space-charge forces effect on the measured momentum spread can be neglected all along the beam path from the entrance of the accelerating structure to the magnetic spectrometer.
2. The effect on the momentum spread of the in- and out-coupling cells and of the leakage fields at the TWS entrance and exit can be neglected.
3. The bunch is much shorter than the wavelength of the accelerating field, so that its effect can be linearized whatever the injection phase.
4. The bunch velocity can be assumed equal to $c$ (speed of light in vacuum), so that the bunch duration and its phase slippage rate respective to the accelerating field are constant all along the structure (the latter being equal to zero if the field phase velocity is $c$).

The less these assumptions are fulfilled, the bigger the error on the reconstructed longitudinal bunch properties. On ARES, the typical charge (a few pC), momentum (35 – 80 MeV/c) and bunch duration (sub-ps rms) at the entrance of TWS2 are such that the assumptions 3 and 4 are largely valid. The assumption 1 is also valid in a large range of parameters, but starts to become less and less valid for short bunches (typically below a few tens of fs rms). The validity of assumption 2 is assessed later in this section.

Under the aforementioned assumptions, the transport of the bunch longitudinal properties can be described by 2*2 matrices in the following way, assuming no coupling between the longitudinal and transverse planes:

$$\Sigma_f = M \Sigma_i M^T, \text{ with } \Sigma_\alpha = \begin{pmatrix} \sigma_{t_\alpha}^2 & \sigma_{pt_\alpha} \\ \sigma_{pt_\alpha} & \sigma_{p_\alpha}^2 \end{pmatrix} \text{ and } M = \begin{pmatrix} R_{55} & R_{56} \\ R_{65} & R_{66} \end{pmatrix} \quad (1)$$

The subscripts $i$ and $f$ respectively refer to the entrance of the accelerating structure which operation condition is varied and the point where the momentum spread is measured. The superscript $T$ refers to the transpose of a matrix. The matrix $M$ is the longitudinal transfer matrix describing the beamline between $i$ and $f$. The transport of the matrix $\Sigma$ leads to the following equations:

$$\begin{cases} \sigma_{t_f}^2 = R_{55}^2 \sigma_{t_i}^2 + 2 R_{55} R_{56} \sigma_{pt_i} + R_{56}^2 \sigma_{p_i}^2 & (2) \\ \sigma_{p_f}^2 = R_{65}^2 \sigma_{t_i}^2 + 2 R_{65} R_{66} \sigma_{pt_i} + R_{66}^2 \sigma_{p_i}^2 & (3) \end{cases}$$

Eq. (2) is not experimentally useful, since it would require to measure the bunch duration at the point $f$ of the beamline to retrieve it at the point $i$. On the opposite, Eq. (3) is of interest since it links the bunch duration $\sigma_{t_i}$ at the point $i$ of the beamline to the momentum spread $\sigma_{p_f}$ measured at the point $f$.

Under the aforementioned assumptions, the bunch momentum spread is invariant in a drift space or when focusing/transport magnets are used. Moreover, on ARES, there are no accelerating structures located downstream of the one used for the measurement (TWS2). This means that the momentum spread already has the value $\sigma_{E_f}$ right at the exit of TWS2, or in other words that only TWS2 contributes to the coefficient $R_{65}$ and $R_{66}$ in Eq. (3). The aforementioned assumptions also lead to $R_{66} = 1$, so that only $R_{65}$ is relevant. For a TWS, the on-axis longitudinal field component seen by an electron under the assumptions 3 and 4 can be written as:

$$E_z(z) = E_m \cos\left(\frac{2\pi f (v_{ph} - c)}{v_{ph} c} z + \varphi_0\right) \quad (4)$$

where $f$ is the TWS resonance frequency, $E_m$ the maximal average gradient, $z$ the position along the structure axis, $c$ the speed of light in vacuum and $v_{ph}$ the phase velocity of the field in the TWS. $\varphi_0$ is the injection phase of the electron bunch in the TWS (180° is the phase of maximum momentum gain rate for $v_{ph} = c$). At ARES, it is desired that $v_{ph}$ is as close as possible to $c$, in order to minimize the phase slippage of the electron bunch. It has to be noted that a purely sinusoidal expression is used in Eq. (4), thus neglecting the spatial harmonics other than the fundamental in the TWS field. This is due to the fact that only the integrated effect throughout TWS2 is relevant for the phase-scan method, and these higher-order spatial harmonics provide no net momentum change when averaged over one period. Defining $\xi(v_{ph}) = \pi f L(v_{ph}-c)/v_{ph}c$, the momentum $p_f$ after TWS2 and $R_{65}$ can be written as:

$$p_f = p_i - \frac{e\widetilde{E_m}L}{c}\cos(\widetilde{\varphi_0}) \qquad (5)$$

$$R_{65} = \frac{2\pi f e \widetilde{E_m} L}{c}\sin(\widetilde{\varphi_0}) \qquad (6)$$

where $L$ is the TWS length, $e$ the fundamental electric charge, $\widetilde{E_m} = E_m \operatorname{sinc}(\xi(v_{ph}))$, $\widetilde{\varphi_0} = \varphi_0 + \xi(v_{ph})$ and $\operatorname{sinc}(x)$ is the function equal to $\sin(x)/x$ for $x \neq 0$ and 1 for $x = 0$.

One can see that, by replacing $E_m$ by $\widetilde{E_m}$ and $\varphi_0$ by $\widetilde{\varphi_0}$, Eqs. (5) and (6) are similar to the ones without phase slippage of the electron bunch. This way of writing is convenient for practical application, because the TWS field amplitude in the model is practically adjusted through experimental measurements of $p_f$. $\widetilde{E_m}$ is therefore the quantity determined in this way, and shown in Table 2, while determining $E_m$ would also require knowing $v_{ph}$. Using $\widetilde{\varphi_0}$ or $\varphi_0$ is strictly equivalent since $\xi(v_{ph})$ is a constant term, thus just implying a global translation of the injection phase scale. Applying the phase-scan method at ARES with TWS2 is therefore independent on the knowledge of its phase velocity under the assumptions made.

All the measurements presented in Sec. 5 are performed by scanning the bunch injection phase $\widetilde{\varphi_0}$ into TWS2 and the momentum spread $\sigma_{p_f}$ is measured by the high-energy spectrometer (see Fig. 1). The reconstructed values of $\sigma_{t_i}$, $\sigma_{pt_i}$ and $\sigma_{p_i}$ from Eq. (3) are thus the ones at TWS2 entrance. This reconstruction requires an input of $n \geq 3$ values of $\sigma_{p_f}$ measured for $n$ values of $\widetilde{\varphi_0}$ ($\equiv R_{65}$). A matrix system of the following form is obtained:

$$Y = AX, \text{ with } Y = \begin{pmatrix} \sigma_{p_{f_1}}^2 \\ \ldots \\ \sigma_{p_{f_n}}^2 \end{pmatrix}, A = \begin{pmatrix} R_{65_1}^2 & 2R_{65_1} & 1 \\ \ldots & \ldots & \ldots \\ R_{65_n}^2 & 2R_{65_n} & 1 \end{pmatrix} \text{ and } X = \begin{pmatrix} \sigma_{t_i}^2 \\ \sigma_{pt_i} \\ \sigma_{p_i}^2 \end{pmatrix} \qquad (7)$$

This matrix system is inverted using a least-square algorithm to obtain the vector $X$ and especially the bunch duration $\sigma_{t_i}$ as follows:

$$X = (A^t A)^{-1} A^t Y \qquad (8)$$

To verify the validity of assumption 2, a comparison with an ASTRA simulation, where the effect of the in- and out-coupling cells and of the leakage fields at TWS2 entrance and exit are included, has been performed in ideal conditions. Namely, the space-charge forces were turned off and it was ensured that the input distribution at TWS2 entrance has no coupling between the longitudinal and transverse planes. This enables to isolate the combined influence of TWS2 in- and out-coupling cells and leakage fields on the phase-scan method accuracy. The bunch rms momentum spread at the high-energy spectrometer has been simulated as a function of the injection phase into TWS2 (see Fig. 2). Eq. (8) has then been applied on the simulated dataset to reconstruct the bunch duration at TWS2 entrance. The reconstructed rms bunch duration (448.23 fs) is extremely close to the input one (448.31 fs), the discrepancy being far below the typical error bars on experimental data and simulations (see Sec. 5). The assumption 2 to neglect the effect of TWS2 in- and out-coupling cells and leakage fields is therefore valid. It is also confirmed that neglecting the spatial harmonics other than the fundamental in TWS2 field has a negligible effect, if any, on the phase-scan method accuracy, since these harmonics are included in ASTRA.

One should note that Eq. (8) does not include any statistical errors (due to experimental jitters) on $\sigma_{p_f}$ and $R_{65}$. To address this, Eq. (8) is combined with a Monte-Carlo algorithm [41]. The starting point is the measured experimental value of $\sigma_{p_f}$, determined as the average of a small number (typically 10 to 30) of measurements, and its rms statistical error, determined as the standard deviation of these measurements. Then, a large number $N$ (typically $N \geq 10^5$) of values of $\sigma_{p_f}$ is randomly generated following a Gaussian distribution with a standard deviation equal to the rms statistical error on $\sigma_{p_f}$. Then, for each of these $N$ values of $\sigma_{p_f}$, one value of $R_{65}$ is randomly drawn following a Gaussian distribution with a standard deviation equal to the rms statistical error on $R_{65}$ (function of the experimental jitters on $\widetilde{E_m}$ and $\widetilde{\varphi_0}$). Applying Eq. (8) on each randomly generated doublet ($\sigma_{p_f}$ ; $R_{65}$) generates a random set of $N$ values for $X$. The final value for $X$ is determined as the average of these $N$ values, and the error bar on it as their standard deviation.

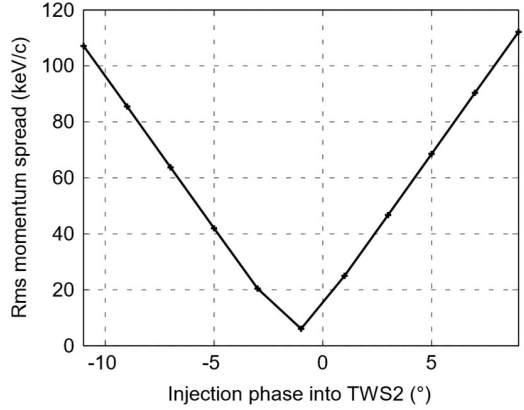

Figure 2: ASTRA simulation of the bunch rms momentum spread at the high-energy spectrometer as a function of the injection phase into TWS2 (0° ≡ max. momentum gain). Conditions: 78.5 MeV/c input momentum; 448.31 fs rms input bunch duration. The application of Eq. (8) on this dataset leads to a reconstructed input bunch duration of 448.23 fs rms.

*3.2. Applicability criterion for the phase scan method*

Qualitatively, a bunch duration threshold below which the phase scan method cannot be applied exists. Namely, when the bunch becomes so short that the spectrometer is unable to resolve at least 3 different values of $\sigma_{p_f}$ when scanning $\widetilde{\varphi_0}$. Mathematically, this translates into the following theoretical applicability criterion for the phase scan method:

$$\sigma_{p_{fmax}} - \sigma_{p_{fmin}} \geq 2R\left(\frac{p_{max} + p_{min}}{2}\right) = 2R\langle p \rangle \qquad (9)$$

with $\sigma_{p_{fmax}}$ and $\sigma_{p_{fmin}}$ being respectively the maximal and minimal value of $\sigma_{p_f}$, $R$ being the relative resolution of the magnetic spectrometer ($10^{-4}$ at ARES) and $p_{max}$ and $p_{min}$ respectively the maximal and minimal beam momentum in the range of phase between $\sigma_{p_{fmax}}$ and $\sigma_{p_{fmin}}$. We therefore approximate here the absolute spectrometer resolution ($R$ times momentum) as equal to the one for the average beam momentum as a function of $\widetilde{\varphi_0}$ in this range, $\langle p \rangle$. It is noteworthy that here, $R$ does not refer to the spectrometer resolution for the measurement of an absolute value of $\sigma_{p_f}$ at a fixed $\widetilde{\varphi_0}$, but to the resolution for the measurement of a variation of $\sigma_{p_f}$ between two values of $\widetilde{\varphi_0}$.

It can be shown that $\sigma^2_{p_{fmax}}$ and $\sigma^2_{p_{fmin}}$ have the following expressions:

$$\sigma^2_{p_{fmax}} = \alpha^2 \sigma^2_{t_i} + 2\alpha|\sigma_{pt_i}| + \sigma^2_{p_i} \qquad (10)$$

$$\sigma^2_{p_{fmin}} = \alpha^2 \sigma^2_{t_i} - 2\alpha|\sigma_{pt_i}| + \sigma^2_{p_i} \; if \; \left|\frac{\sigma_{pt_i}}{\alpha\sigma^2_{t_i}}\right| \geq 1 \qquad (11)$$

$$\sigma^2_{p_{fmin}} = \sigma^2_{p_i} - \frac{\sigma^2_{pt_i}}{\sigma^2_{t_i}} \; if \; \left|\frac{\sigma_{pt_i}}{\alpha\sigma^2_{t_i}}\right| \leq 1 \qquad (12)$$

with $\alpha = 2\pi f e\widetilde{E_m}L/c$. From Eqs. (9), (10), (11) and (12), the criterion for applicability of the phase scan method can be verified for any input bunch and TWS2 properties.

It is noteworthy that the applicability criterion above is derived in ideal conditions, and therefore give the ultimate limit of the phase scan method. It does not give the accuracy or resolution of the phase scan method. Two important effects exist that prevent to measure a bunch duration as short as the applicability criterion would allow with the phase scan method. First, the criterion is defined such that only 3 values of $\sigma_{p_f}$ can be resolved, which is the very minimum to apply the phase scan method. Under these conditions, the retrieved $\sigma_{t_i}$ is sensitive to jitters and errors on a single value of $\sigma_{p_f}$. This can be significantly mitigated by recording more values of $\sigma_{p_f}$ (typically around 10), so that in practice $2R\langle p \rangle$ in Eq. (9) should be replaced by a higher number (typically $9R\langle p \rangle$ to be conservative). Second, for ultrashort bunches the space-charge forces induce a not constant bunch duration throughout TWS2 and also modify the value of $\sigma_{p_f}$ between TWS2 entrance and the high-energy spectrometer. As a result, the retrieved value of $\sigma_{t_i}$ is also modified, meaning in practice a worse resolution than the theoretical applicability criterion would allow. In other words, this means that the assumptions 1 and 4 of Sec. 3.1 are not anymore fulfilled.

## 4. Influence and determination of the TWS1 field phase velocity

The primary goal of ARES is to generate ultrashort electron bunches. One of the schemes that can be applied to this aim is to compress the bunch via the velocity bunching process [36] in TWS1. This process is especially sensitive to the bunch input momentum (at TWS1 entrance) and to the field phase velocity $v_{ph}$ in TWS1. These two parameters significantly influence the way the not ultra-relativistic electron bunch from the ARES gun (typically 3.5 to 4 MeV/c) is compressed in time via the velocity bunching process. On the one hand, the input bunch momentum can be measured with an uncertainty better than a few percent with the low-energy spectrometer (see Fig. 1). On the other hand, to the best of our knowledge, there is no conventional method to precisely measure $v_{ph}$ once a TWS is installed on an accelerator. Nevertheless, in order to be able to compare the bunch duration measured at TWS2 entrance via the phase-scan method with the prediction from ASTRA simulations, it is essential to precisely know $v_{ph}$ for TWS1. Fig. 3 illustrates this by comparing the TWS1 compression curve, namely the bunch duration after TWS1 as a function of the injection phase into it, simulated with ASTRA for several phase velocities. It shows that a deviation of +/- 1‰ respective to $c$ (the design value for the TWS at ARES) leads to a significant distortion of the TWS1 compression curve, especially the injection phase leading to maximal compression is shifted by around +/- 5°. The achieved minimal bunch duration remains however very similar for the different phase velocities.

For TWS2, the phase velocity has no significant influence on the electron bunch duration since it is already ultra-relativistic after TWS1 (at least 30 MeV/c and up to 79 MeV/c) and therefore almost frozen. In addition, as shown in Sec. 3.1, the knowledge of $v_{ph}$ in TWS2 is not required to apply the phase-scan method.

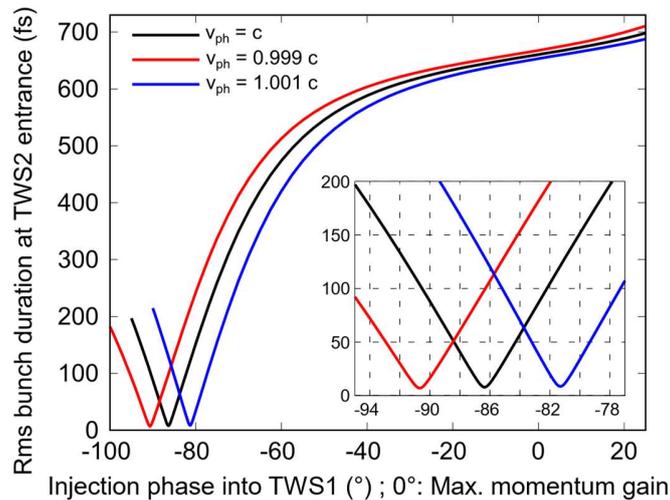

**Figure 3: Simulated rms bunch duration after TWS1 as a function of the injection phase into it for several phase velocities $v_{ph}$. The inset shows a zoom around the maximal compression. Conditions: 1 pC charge; 320 μm transverse diameter of laser pulse driving the gun; 69.5 MV/m gun peak accelerating field (3.64 MeV/c output bunch); See Table 2.**

To experimentally determine a TWS phase velocity, we propose a method taking advantage of a not ultra-relativistic input electron bunch, e.g. delivered by an RF-gun. For such a bunch, the curve of the momentum after a TWS as a function of the injection phase exhibits a pattern with 2 maxima and a saddle point in between. The shape of this curve, all other parameters being fixed, strongly depends on the TWS phase velocity, as it is visible in Fig. 4.a. Several features of this curve can thus be used to experimentally determine the TWS phase velocity, e.g. the momentum value at the saddle point or the injection phase gap $\Delta\Phi$ between the maximum and the saddle point. Here we propose to use $\Delta\Phi$, because it is not affected by a potential miscalibration of the high-energy spectrometer. The calibration curve $\Delta\Phi(v_{ph})$, generated through ASTRA simulations, for both TWS1 and TWS2 is given in Fig. 4.b. One has to note that the calibration curve remains very similar for TWS1 and TWS2 despite their different average gradients. They would be almost superimposed and therefore only one is displayed on Fig. 4.b for visibility reason. This shows that a limited change of the TWS average gradient (at least 1.1 MV/m for a 3.85 MeV/c input bunch) has a negligible effect on $\Delta\Phi(v_{ph})$ for the range considered at ARES (+/- 0.0015 $c$ around $c$). The quadratic fit on Fig. 4.b has the following expression, with $v_{ph}$ in units of $c$:

$$\Delta\Phi(°) = -90880.95 v_{ph}^2 + 176334.65 v_{ph} - 85364.33 \qquad (14)$$

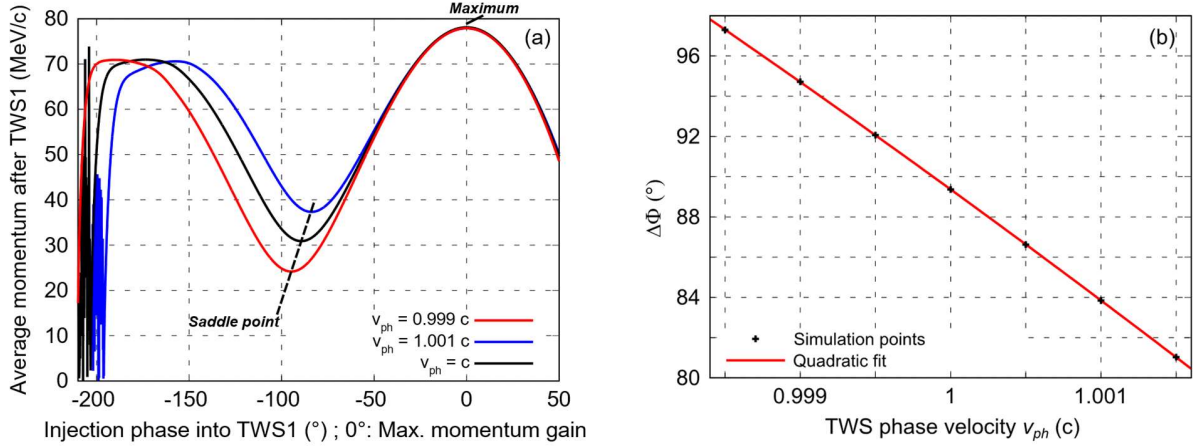

Figure 4: (a): Simulated average momentum after TWS1 as a function of the injection phase into it for several phase velocities $v_{ph}$. (b): Phase gap $\Delta\Phi$ between the maximum momentum and saddle point as a function of $v_{ph}$. Conditions: Single particle simulation; 74 MV/m gun peak accelerating field (3.85 MeV/c output particle); See Table 2.

Measurements of $\Delta\Phi$ have been performed at ARES for both TWS, the other TWS being switched off. Fig. 5 shows the raw data used to determine $\Delta\Phi$ (average momentum versus injection phase). The values of $\Delta\Phi$ displayed in Fig. 5 have been determined through two parabolic fits of the experimental data, one around the maximum and one around the saddle point, optimized via a Monte-Carlo algorithm [41] based on the experimental error bars. Using the calibration of Fig. 4.b, they lead to $v_{ph}$(TWS1) = (1.00040 +/- 0.00009) $c$ ($\Delta\Phi$ = 87.2° +/- 0.5°) and $v_{ph}$(TWS2) = (1.00063 +/- 0.00009) $c$ ($\Delta\Phi$ = 85.9° +/- 0.5°). On Fig. 5, ASTRA simulations where these experimentally determined $v_{ph}$ are used are also shown for comparison. Due to different conventions, the sign of the injection phase into an accelerating structure is inverted between ASTRA and the experiment on ARES, namely a positive phase in ASTRA is negative in experiment on ARES and vice-versa. For the purpose of comparison, the sign of the phase in ASTRA is inverted in Fig. 5. This will also be the case later for Figs. 9, 10 and 11 in Sec. 5.2.

A last important step is to quantify the effect of the typical uncertainty on the input bunch momentum at ARES on the measurement procedure described above. This is required, because a not ultra-relativistic input bunch momentum also has a significant influence on the phase slippage in the TWS. An error on its determination therefore comes with an error on the determination of the calibration curve of Eq. (14), since it is determined for a fixed value of the input bunch momentum, and subsequently on the determination of $v_{ph}$. To this aim, it has been simulated which error on the input bunch momentum (measured at 3.85 MeV/c) would be necessary to retrieve the same $\Delta\Phi$ than on Fig. 5 assuming $v_{ph}$(TWS1) = $c$ (the design value). Fig. 6 shows that an input bunch momentum of around 3.12 MeV/c, translating into an error of 23.4%, would be required to reach the measured $\Delta\Phi$ = 87.2° if $v_{ph}$(TWS1) was equal to $c$. This is much higher than the uncertainty on the bunch momentum measurement at ARES, which is typically of a few percent at maximum, corresponding to around 0.1 MeV/c. Besides, this maximal error of 0.1 MeV/c translates into an error of around 0.25° on $\Delta\Phi$, which is significantly lower than the uncertainty on the experimental measurement of $\Delta\Phi$ due to machine jitters (0.5°). This demonstrates that the uncertainty on the input bunch momentum would only marginally modify the determined value of $v_{ph}$. To take this into account, its contribution to the uncertainty on $\Delta\Phi$ (0.25°) is quadratically added to the one from the machine jitters (0.5°), resulting in an overall uncertainty of 0.56°. This increases the uncertainty on $v_{ph}$ from 0.00009 $c$ to 0.0001 $c$.

As mentioned before, the knowledge of $v_{ph}$(TWS2) is not required for the study presented in this paper. It is however of interest to be measured, for comparison with $v_{ph}$(TWS1). Indeed, although the two TWS at ARES are based on the same design, it is visible that their $v_{ph}$ are different, namely they do not overlap within the error bars. It shows that $v_{ph}$ is specific to a single TWS when installed on an accelerator, and that it has to be characterized separately for each of them. The method described in this section is suited for this and can be applied on any accelerator, with a precision down to sub-permille variation of $v_{ph}$ respective to $c$. The condition for application is that a not ultra-relativistic beam (typically ≤ 5 MeV/c) of precisely known momentum (uncertainty better than a few percent) can be delivered at the entrance of the TWS and a downstream momentum measurement is available.

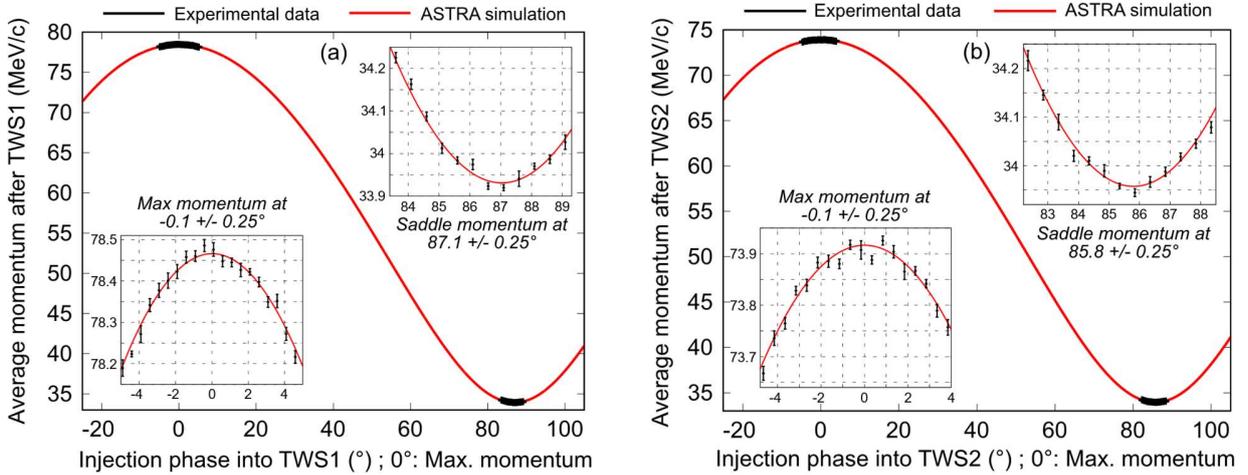

**Figure 5:** Raw data for experimental determination of TWS 1 (a) and TWS2 (b) phase velocities (average momentum as a function of injection phase). The insets show zooms around the maximum and saddle momentum phase ranges. Conditions: Mo cathode; 0.9 pC (a) and 0.25 pC (b) charge; 320 µm transverse diameter of laser pulse driving the gun; 74 MV/m gun peak accelerating field (3.85 MeV/c output particle); See Table 2.

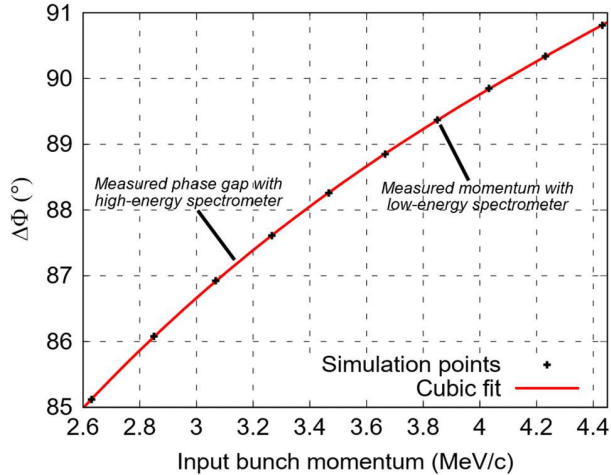

**Figure 6:** Phase gap ΔΦ between the maximum momentum and saddle point at the exit of TWS1 as a function of the input bunch momentum. Conditions: Single particle simulation; $v_{ph}$(TWS1) assumed equal to $c$ (the design value); See Table 2.

## 5. Bunch duration measurements at ARES using the phase-scan method

*5.1. Comparison of Mo cathode and Cs$_2$Te cathode*

Two different types of cathodes are used as electron source in the ARES RF-gun: Mo (metal) and Cs$_2$Te (semi-conductor). From the literature, it is known that Cs$_2$Te has a longer response time than Mo. Direct measurements of Cs$_2$Te cathode response times in RF-guns have been performed using electron micro-bunches with adjustable time separation generated by laser interferometry [42,43]. A value around 370 fs has been measured using a zero-phasing technique as diagnostic on the LUCX facility at KEK [42], and more recently values between 180 fs and 250 fs (for different cathodes) have been measured on the PITZ facility at DESY using a TDS [43].

The cathode laser used at ARES to drive the RF-gun has a duration of 175 fs rms, which is slightly shorter than the shortest $Cs_2Te$ response time reported in literature (180 fs) and is expected to be significantly longer than the Mo response time. Experimental measurements of the bunch duration at ARES in exactly the same conditions for a Mo and a $Cs_2Te$ cathode should therefore give different values, due to the different response times. A comparison with ASTRA simulations with a varying response time can then be used to estimate the response time of the $Cs_2Te$ cathode mounted in the ARES RF-gun. This estimated value will then be used later in the paper when comparing experimental measurements made with $Cs_2Te$ cathodes with simulations.

A bunch duration measurement using the phase-scan method has been performed in identical conditions for a Mo and a $Cs_2Te$ cathode. The raw data used for the reconstruction of the bunch duration (momentum spread against injection phase into TWS2) are shown in Fig. 7. It is visible that the momentum spread for the $Cs_2Te$ cathode increases faster with phase than for the Mo cathode when moving away from the minimum, which is due to a longer bunch duration at TWS2 entrance resulting from the $Cs_2Te$ longer response time. An analysis of the data in Fig. 7 with the model of Sec. 3.1 gives at TWS2 entrance a bunch duration of (567.7 +/- 3.4) fs rms for the Mo cathode and (585.6 +/- 4.1) fs rms for the $Cs_2Te$ cathode.

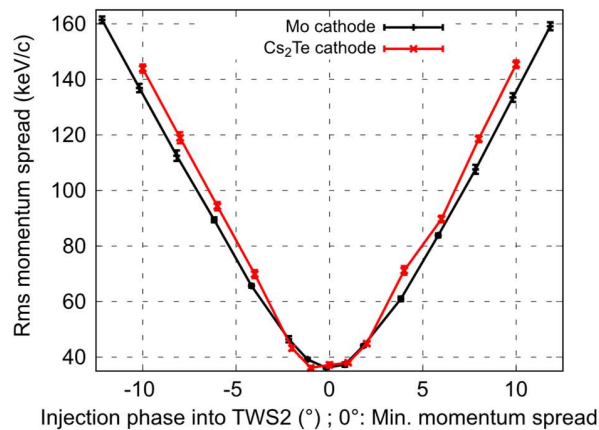

**Figure 7:** Raw data for bunch duration determination at TWS2 entrance (rms momentum spread as a function of injection phase) for Mo and $Cs_2Te$ cathodes in identical conditions. Conditions: (1.07 +/- 0.015) pC charge; 320 µm transverse diameter of laser pulse driving the gun; 70.6 MV/m gun peak accelerating field (3.69 MeV/c output momentum); Injection phases into RF-gun and TWS1 adjusted to maximize the momentum; See Table 2.

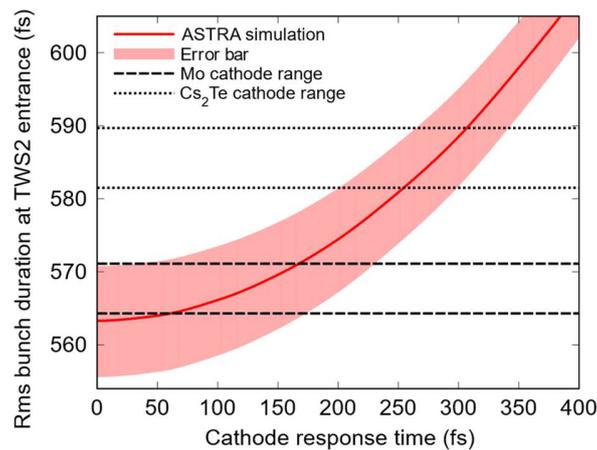

**Figure 8:** Rms bunch duration at TWS2 entrance as a function of the cathode response time simulated by ASTRA compared with the experimental values measured for a Mo and a $Cs_2Te$ cathode with their uncertainty ranges. The error bar on the ASTRA simulation has been computed by including the jitter on the bunch charge, an uncertainty of +/- 0.25 A on the current injected into the focusing solenoid located after the RF-gun and the uncertainty on TWS1 phase velocity. Conditions: See Figure 7 caption.

A comparison of the aforementioned bunch durations with those obtained in ASTRA simulations as a function of the cathode response time is shown in Fig. 8. From Fig. 8, it is visible that the bunch duration measured for the Mo cathode is compatible with a very short response time (we will assume it equal to 0 in the rest of the paper), while the one measured for the $Cs_2Te$ cathode points towards a higher response time in the range between 200 and 300 fs. A detailed Monte-Carlo analysis of the overlap between the simulation curve and the range for the experimentally measured duration gives an estimated response time of (271.4 +/- 32.4) fs (at one standard deviation) for the $Cs_2Te$ cathode. This estimated value is within the range coming from direct experimental measurements [42,43], and is therefore compatible with them. It will be used for comparison with simulations in Sec. 5.2 when a $Cs_2Te$ cathode is mounted in the ARES RF-gun.

*5.2. Systematic bunch duration measurements with the phase-scan method at ARES*

The bunch duration at TWS2 entrance has been measured at ARES as a function of three parameters having a significant influence on it: the RF-gun field amplitude (Fig. 9.a), the bunch charge (Fig. 9.b) and the injection phase into TWS1 (Fig. 9.c). Aside of each bunch duration curve, a selection (for visibility reason) of the raw data used for the measurement (rms momentum spread as a function of the injection phase into TWS2) is also displayed. The cathode in use when acquiring the data for Fig. 9.c was producing a large amount of dark current through field emission [44]. To reduce it, a 1 mm diameter collimator was used in the gun region, before TWS1 (see Fig. 1).

The experimental results are compared with the prediction from ASTRA simulations, in order to evaluate our understanding of the ARES operation on the aspect of bunch duration. Note that for this purpose, the TWS1 phase velocity determined in Sec. 4 is used. For Figs. 9.a to 9.c, the error bar on the ASTRA simulation includes the jitter on the bunch charge, an uncertainty of +/- 0.25 A on the current injected into the focusing solenoid located after the RF-gun and the uncertainty on the TWS1 phase velocity. For Fig. 9.c, since a $Cs_2Te$ cathode and a 1 mm diameter collimator were used, the uncertainties on the cathode response time and the bunch charge after the collimator (measured with a resonator [45]) are also included.

The experimental results in Figs. 9.a to 9.c are in very good agreement with the ASTRA simulations, almost all the points matching within the error bars. This demonstrates a very good understanding of the ARES operation on the bunch duration aspect for a large range of parameters. The preparatory experimental determination of the TWS1 phase velocity (Sec. 4) is especially important to compare Fig. 9.c with ASTRA simulations. This is illustrated by Fig. 10, where the experimental data of Fig. 9.c are compared with ASTRA simulations where a TWS1 phase velocity equal to *c* (the design value) is used instead of the measured 1.00040 *c* used in Fig. 9.c. It is clearly visible that, even for this small change of the phase velocity, a significant discrepancy between the ASTRA simulation and the experimental data, not covered by the error bars, would appear when approaching and overcoming the TWS1 phase for maximal compression.

Fig. 11 shows a zoom of Fig. 9.c around the compression maximum, where a curve is superimposed to evaluate if the applicability criterion derived in Sec. 3.2 is fulfilled. For this purpose, the variable $(\sigma_{p_{fmax}} - \sigma_{p_{fmin}})/(R\langle p \rangle)$ is used. As visible in Fig. 11, it is greater than 2 for all the experimental data, meaning that the theoretical applicability criterion from Sec. 3.2 is fulfilled. It is not fulfilled only in a small range between -84.1° and -84.55°, where no experimental data has been acquired. However, it is also visible that for the experimental data at -83.8° and -84.8°, the variable $(\sigma_{p_{fmax}} - \sigma_{p_{fmin}})/(R\langle p \rangle)$ is only around 5. As explained in Sec. 3.2, this tends to increase the uncertainty of the phase scan method, because only a small number of data points can then be used as input for the phase scan method. This is especially visible for the point at -83.8° where the uncertainty on the retrieved bunch duration approaches 50%.

One can also remark on Fig. 11 that the 3 points at -83.8°, -84.8° and -85.8° do not follow the shape of the simulation curve and are rather aligned along a flat line around 20 fs rms, which is a sign of a practical resolution limit of the phase scan method under the ARES operation conditions. As explained in Sec. 3.2, this is very likely due to the space-charge forces. They become not negligible for such short bunches and therefore imply a not constant bunch duration throughout TWS2 and a change of the momentum spread between TWS2 and the high-energy spectrometer, thus affecting the bunch duration retrieved from the phase scan method. This value of 20 fs rms is, to the best of our knowledge, around 4 times shorter than what has been previously experimentally demonstrated based on RF-phasing techniques with a single RF structure [23]. It could be slightly improved, within the limits of the applicability criterion discussed in Sec. 3.2, by reducing the space-charge forces effect. This could be achieved by reducing the bunch charge and/or the distance between TWS2 and the high-energy spectrometer. In general, but not applicable to ARES, using several TWS would also improve the resolution by allowing inducing more momentum spread variation.

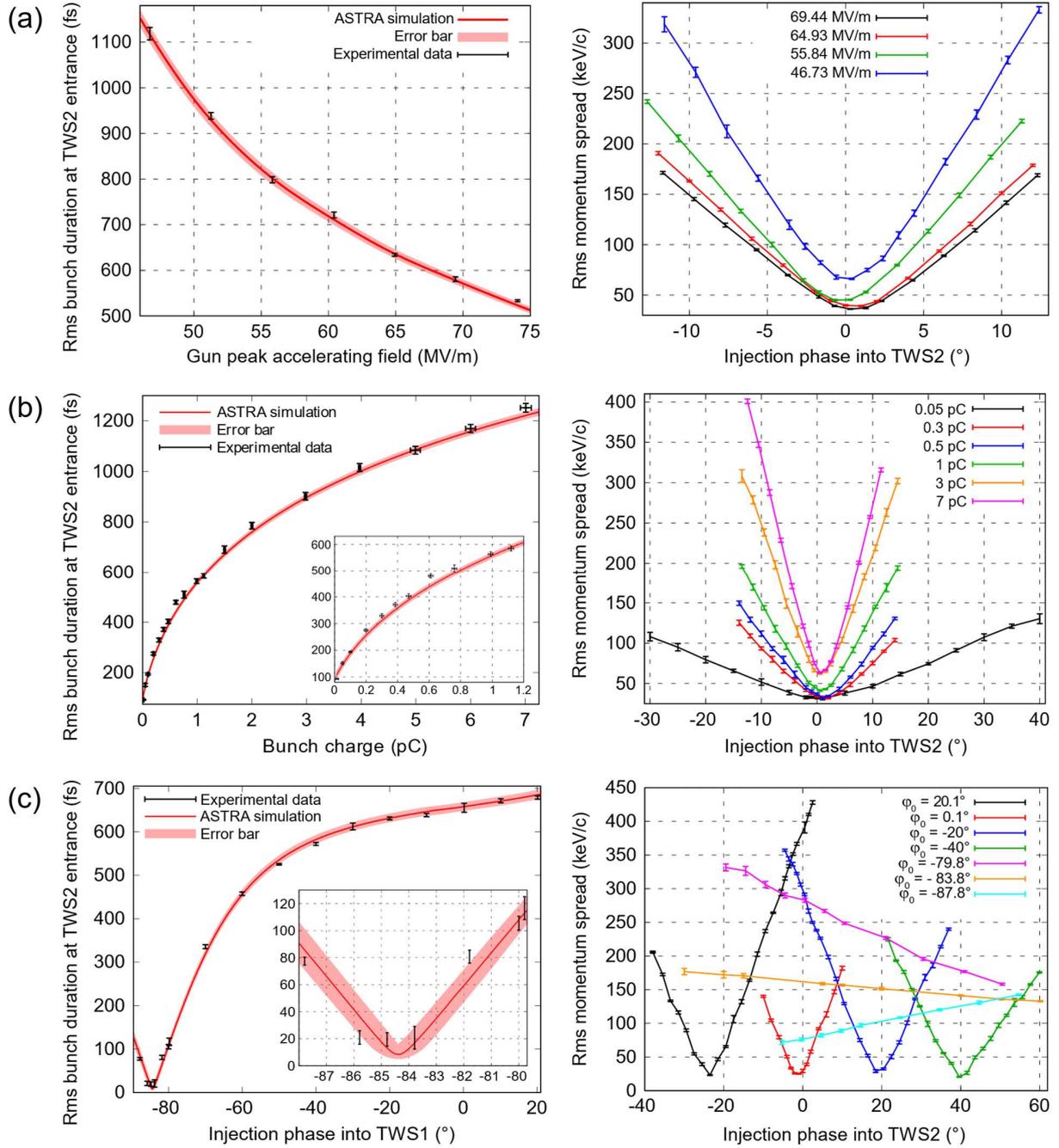

**Figure 9:** Rms bunch duration at TWS2 entrance as a function of the RF-gun peak accelerating field (a), the bunch charge (b) and the injection phase into TWS1 (c). 0° ≡ max. momentum gain. Aside of each duration curve, a selection of the raw data used for the measurements is displayed. Conditions: Mo (a,b) and $Cs_2Te$ (c) cathode; 320 µm (a,c) and 800 µm (b) transverse diameter of laser pulse driving the RF-gun; 64.9 MV/m (b) and 69.5 MV/m (c) gun peak accelerating field (resp. 3.43 and 3.64 MeV/c output momentum); (1.08 to 1.05 +/- 0.02) pC depending on point (a) and (1.27 +/- 0.02) pC before collimator and (1.01 +/- 0.02) pC after collimator (c); Injection phases into RF-gun adjusted to maximize the momentum (except for the points at 46.7 MV/m (resp. 51.3 MV/m) in (a) where it is fixed to +6.7° (resp. +11.7°) above the zero-crossing field); Injection phase into TWS1 adjusted to maximize the momentum gain (a,b); See Table 2. Note: For (b), the error bars on the charge for the 3 highest points are larger because of a different setting of the Faraday Cup. The charge for the lowest point (0.015 pC) has not been directly measured with the Faraday Cup but extrapolated from the cathode laser settings.

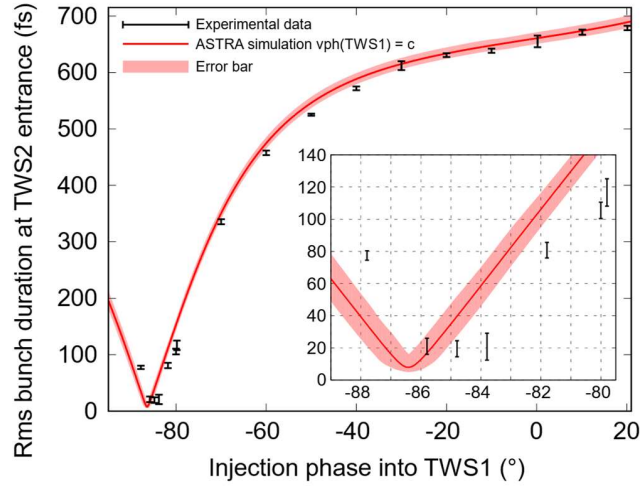

Figure 10: Rms bunch duration at TWS2 entrance as a function of the injection phase into TWS1 (0° ≡ max. momentum gain). The phase velocity of TWS1 is here set to *c* (the design value) in the ASTRA simulation, instead of the experimentally measured value (1.0004 *c*) as in Fig. 9.c. Conditions: See Fig. 9.c.

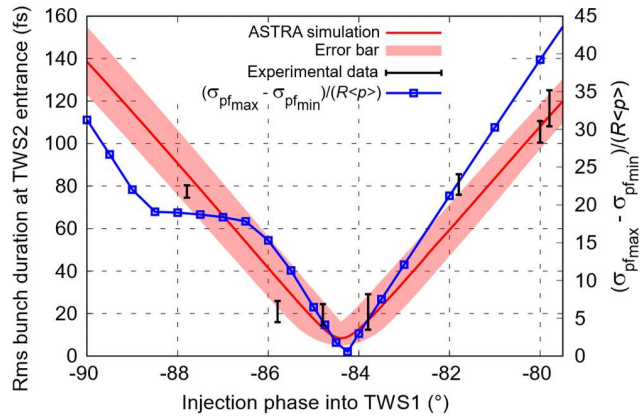

Figure 11: Rms bunch duration at TWS2 entrance and $\left(\sigma_{p_{fmax}} - \sigma_{p_{fmin}}\right)/(R\langle p\rangle)$ as a function of the injection phase into TWS1 (0° ≡ max. momentum gain). The variable $\left(\sigma_{p_{fmax}} - \sigma_{p_{fmin}}\right)/(R\langle p\rangle)$ is used to check the applicability of the phase scan method (see Sec. 3.2). Conditions: See Fig. 9.c.

## 6. Conclusion and outlook

The commissioning phase leading to the first characterization of the duration of the electron bunches generated by the ARES linac at DESY has been presented. An RF-phasing technique, the phase-scan method, has been applied with a 4.092 m long S-band travelling wave accelerating structure for this purpose. The sensitivity of the method allowed highlighting different response times for the Mo and $Cs_2Te$ cathodes mounted in the ARES RF-gun, the value of around 270 fs for the $Cs_2Te$ cathode estimated from comparison with ASTRA simulations being consistent with direct measurements reported in the literature. The overall very good agreement between the experimental measurements and the predictions from ASTRA simulations demonstrates that the ARES operation is well understood on the bunch duration aspect for a large range of operation parameters (charge, gun peak field and injection phase into TWS1). An important requirement for that is the preliminary determination of the phase velocity in TWS1. To this aim, we proposed a simple beam-based method precise down to sub-permille variation respective to *c*. This method is applicable on any accelerator at the condition that a not ultra-relativistic beam (typically ≤ 5 MeV/c) of precisely known momentum (uncertainty better than a few percent) can be delivered at the entrance of a TWS and a downstream momentum measurement is available.

The shortest bunch duration measured on ARES with the phase scan method after compression via velocity bunching in TWS1 is around 20 fs rms, which is very likely limited by the influence of the space-charge forces. This represents the first experimental demonstration of the ARES linac ability to generate ultrashort electron bunches. It is therefore a strong basis for future demonstrations towards the ARES objective of even shorter bunches (fs to sub-fs scale), which will require a dedicated diagnostic to be resolved. Two PolariX-TDS [31-33] are expected to be commissioned at ARES in the second half of 2023 to fulfill this purpose. The phase-scan method, now routinely used at ARES, will be an important benchmark tool during the commissioning phase of the PolariX-TDS and will also continue to serve as additional diagnostics thereafter.

The capability of ARES to generate ultrashort electron bunches in the range around 100 MeV combined with the availability of conventional diagnostics will be of primary importance for the TWAC project [46], which aims to develop THz-driven structures for the purpose of acceleration and compression of electron bunches down to tens of fs rms. The ARES linac will serve as a test bench for the development of advanced compact duration diagnostics within the framework of this project.

**Acknowledgments**

The authors acknowledge support from DESY (Hamburg, Germany), a member of the Helmholtz Association HGF. The authors would like to thank the technical groups at DESY for the installation, development and regular maintenance of the ARES linac. The authors also thank U. Dorda and B. Marchetti for their leading work during the installation, conditioning and early commissioning phases of the ARES linac. The authors thank K. Flöttmann for his careful reading of the manuscript and his suggestions to improve it. Funding has been received from the European Union's Horizon Europe research and innovation programme (EIC Pathfinder) under grant agreement n. 101046504. Views and opinions expressed are those of the authors only and do not necessarily reflect those of the European Union or EISMEA. Neither the European Union nor the granting authority can be held responsible for them.